\documentclass{article}
\usepackage{spconf,amsmath,graphicx, hyperref}
\usepackage{booktabs} 
\usepackage[symbol]{footmisc}
\usepackage{multicol}
\usepackage{lipsum}

\title{Leveraging Complementary Attention maps in vision transformers for OCT image analysis}

\name{
\begin{tabular}{@{}c@{\hspace{1cm}}c@{\hspace{1cm}}c@{}}
Haz Sameen Shahgir $^{\star}$ &
Tanjeem Azwad Zaman $^{\star}$ &
Khondker Salman Sayeed \\
Md. Asif Haider & 
\textit{Sheikh Saifur Rahman Jony} & 
\textit{M. Sohel Rahman} \\
\end{tabular}
}

\address{Department of Computer Science and Engineering\\
	Bangladesh University of Engineering and Technology\\
    $^{\star}$ Equal Contributions; listed alphabetically.}

\begin{document}
%
\maketitle
\begin{abstract}

Optical Coherence Tomography (OCT) scan yields all possible cross-section images of a retina for detecting biomarkers linked to optical defects. Due to the high volume of data generated, an automated and reliable biomarker detection pipeline is necessary as a primary screening stage.

We outline our new state-of-the-art pipeline for identifying biomarkers from OCT scans. In collaboration with trained ophthalmologists, we identify local and global structures in biomarkers. Through a comprehensive and systematic review of existing vision architectures, we evaluate different convolution and attention mechanisms for biomarker detection. 
We find that MaxViT, a hybrid vision transformer combining convolution layers with strided attention, is better suited for local feature detection, while EVA-02, a standard vision transformer leveraging pure attention and large-scale knowledge distillation, excels at capturing global features.
We ensemble the predictions of both models to achieve first place in the IEEE Video and Image Processing Cup 2023 competition on OCT biomarker detection, achieving a patient-wise F1 score of 0.8527 in the final phase of the competition, scoring 3.8\% higher than the next best solution. Finally, we used knowledge distillation to train a single MaxViT to outperform our ensemble at a fraction of the computation cost.

\vspace{.45em}
\end{abstract}

\vspace{-.5em}

\section{Introduction}
\label{sec:intro}

\vspace{-.5em}
Optical Coherence Tomography (OCT) has revolutionized ophthalmology by providing detailed cross-sectional imaging of retinal structures. The high volume of images generated during OCT scanning---typically hundreds of cross-sections per patient---necessitates automated analysis for practical clinical deployment. While early work by \cite{tranferOCT} demonstrated the potential of automated approaches through transfer learning, and subsequent studies \cite{retinaldisease} refined these techniques, the simultaneous detection of multiple biomarkers remains challenging due to their diverse manifestations.

Through collaboration with clinical experts, we identified that OCT biomarkers exhibit distinctly different spatial characteristics. Some biomarkers, such as Intraretinal Hyperreflective Foci (IRHRF), appear as localized anomalies, while others, like Partially Attached Vitreous Face (PAVF), can only be identified by examining global retinal structure. This fundamental insight suggests that a single architectural approach may be suboptimal for comprehensive biomarker detection.

This observation motivated our systematic study of vision architectures, evaluating their effectiveness in detecting both local and global features in OCT scans. Our investigation revealed that different architectural paradigms excel at different scales: MaxViT's combination of convolution layers and strided attention proved particularly effective for local feature detection, while EVA-02's standard $O(n^2)$ attention mechanism demonstrated superiority in capturing global patterns.

Our key contributions are:
\begin{itemize}
\vspace{-.7em}
    \item A systematic evaluation of vision architectures for OCT analysis, revealing the importance of architectural choices for different types of biomarkers
    \vspace{-.7em}
    \item Classification of OCT biomarkers based on their spatial characteristics, supported by clinical expertise, leading to targeted architectural solutions
    \vspace{-.7em}
    \item Development of an efficient pipeline combining specialized models for local and global feature detection, through ensembling for maximum accuracy and knowledge distillation for computational efficiency
\end{itemize}
\vspace{-.7em}
This approach achieved state-of-the-art performance in the IEEE Video and Image Processing Cup 2023 competition on OCT biomarker detection, demonstrating the practical value of our methodology. Moreover, our final distilled model maintains high accuracy while being computationally efficient enough to handle the high throughput demanded by clinical OCT scanning.
\begin{figure*}[htbp]
    \centering
    \includegraphics[width=1\linewidth]{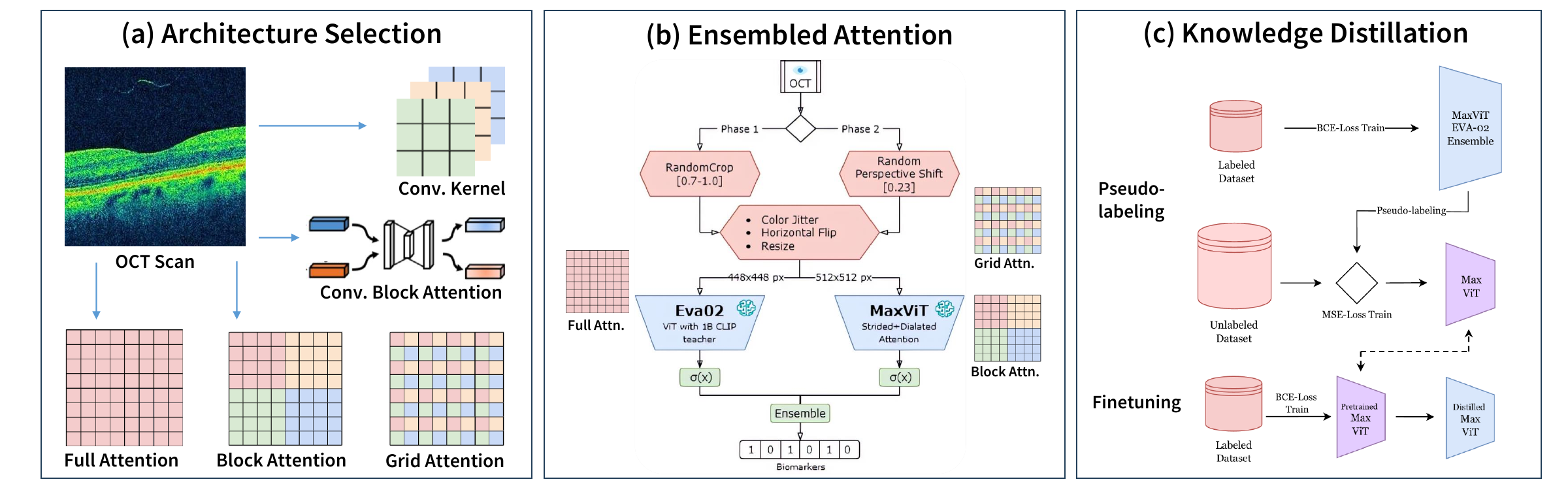}
    \caption{(a) Optimal architecture selection through comprehensive and systematic evaluation. (b) Ensembled MaxViT and EVA02 for local and global biomarker detection respectively. (c) Knowledge distillation via pseudo-labeling.}
    \label{fig:my_label}
\end{figure*}

\vspace{-.7em}
\section{Related Works}
\label{sec:related_works}
\vspace{-.7em}

Recent literature in ophthalmology has shown various approaches to automated OCT analysis. Work by Kermany et al. established early success in OCT classification using transfer learning, achieving 98\% accuracy in identifying conditions like CNV, DME, and DRUSEN \cite{tranferOCT}. Building on this, researchers explored binary CNN classifiers with feature extractors like VGG16 and InceptionV3, achieving 98.7\% accuracy in disease classification \cite{retinaldisease}.

Moving towards more sophisticated architectures, subsequent work introduced a specialized CNN architecture for distinguishing retinal layer degenerations, achieving near-perfect accuracy rates of 99.8\% \cite{opticnet}. Further advances came through a joint-attention-network mechanism, achieving 100\% accuracy on the Srinivasan2014 dataset and 92.40\% on the OCT2017 dataset \cite{jointattentionOCT}. Most recently, RASP-Net focused on identifying and quantifying 11 chorioretinal biomarkers, achieving a mean balanced accuracy of 0.916 and introducing 3D macular profile reconstruction \cite{segquantOCT}.
Other works have explored advanced preprocessing steps for OCT images in conjunction with standard convolutional models \cite{eeepaper}.

While these approaches showcase increasing sophistication in OCT analysis, they largely rely on standard computer vision architectures adapted from other domains. OCT scans contain unique structural patterns at varying scales that may benefit from more specialized architectural considerations.

\vspace{-.8em}
\section{Methodology}
\label{sec:method}
\vspace{-.8em}

\subsection{Dataset}
\vspace{-.3em}
For training, we utilized OLIVES \cite{prabhushankarolives2022}, a rich dataset encompassing 9408 labeled image-biomarker pairs collected from 96 patients over a period of 100 weeks and an additional 78185 unlabeled OCT images, each accompanied by clinical labels. We used an 80:20 train-validation split for pilot experiments. For our final ensemble model we used cross-validation for full utilization of the entire training dataset.
We evaluated our solution on two different test datasets as follows. Aimed at evaluating generalization capabilities, test dataset 1 comprised 3871 images from 40 different patients. To evaluate personalization capabilities, we used test dataset 2 consisting of 250 images collected from 167 new patients. We refer to these datasets as \textit{Testset-1} and \textit{Testset-2} respectively.

Each OCT scan segment had labels to denote the presence or absence of 6 biomarkers, namely Intraretinal Hyperreflective Foci (IRHRF), Partially Attached Vitreous Face (PAVF), Fully Attached Vitreous Face (FAVF), Intraretinal Fluid (IRF), Diffuse Retinal Thickening or Diabetic Macular Edema (DRT/DME) and Vitreous Debris (VD). Depending on the spatial extent, IRHRF and IRF can be loosely grouped as \textit{local} features, meaning they could be detected by looking at just a subsection of the image. On the other hand, PAVF, FAVF, and VD are 
 \textit{global} features, with DRT/DME falling in between.

\vspace{-1.1em}
\subsection{Models Considered}
\vspace{-.4em}
\label{ssec:model}
 
 We considered multiple variants of ResNet \cite{he2016deep} models and Inception \cite{szegedy2017inception} models (collectively referred to as Convolution-based Models henceforth). 
Inspired by \cite{CBAM_OCT}, we added Convolutional Block Attention Modules (CBAM) \cite{CBAM} to InceptionResnetV2 (referred to as IRV2\_CBAM for brevity). We added three such CBAMs after the Stem, Reduction A, and Reduction B modules of InceptionResnetV2. The improved performance of IRV2\_CBAM (to be presented in Section \ref{sec:results}) inspired us to move to vision transformer models, including ViT \cite{dosovitskiy2020image},  MaxViT \cite{tu2022maxvit}, and EVA-02 \cite{EVA02}.

Our early tests indicated an important role for image dimensions when detecting biomarkers. We consulted with multiple trained ophthalmologists and they confirmed that downsizing images to a resolution of 224x224 pixels might have made it harder to identify these biomarkers. 
As such, we focused on models pre-trained on larger images. ViT \cite{dosovitskiy2020image},  We use the base configurations of MaxViT \cite{tu2022maxvit} and EVA-02 \cite{EVA02} which support image resolutions of  $384 \times 384$, $512 \times 512$ and $448 \times 448$ respectively.

\vspace{-.8em}
\subsection{Ensembling MaxViT and EVA-02}
\vspace{-.4em}

The complementary strengths of MaxViT (for local biomarkers) and EVA-02 (for global biomarkers) naturally imply that ensembling their outputs would improve upon their individual performance across all biomarkers. One straightforward way to implement this is by using MaxViT to detect local biomarkers while entirely ignoring its predictions for global biomarkers, and vice versa for EVA-02 (disregarding its local biomarker predictions and using it only for global biomarker prediction). We also apply a finer-grained ensembling scheme, where we average both model's output probabilities. Fig. \ref{fig:my_label} presents a schematic overview of our overall pipeline. We will refer to this (finer-grained) ensemble as MaxViT-EVA02. 

\vspace{-.8em}
\subsection{Knowledge Distillation}
\vspace{-.4em}
We used our MaxViT-EVA02 to pseudo-label the unlabeled data. Using these pseudo-labels, we pre-trained a MaxViT model from scratch with a Mean Squared Error (MSE) loss and subsequently fine-tuned it on the labeled data. This pipeline resulted in substantial performance improvements.

\vspace{-.7em}
\subsection{Evaluation Metrics}
\vspace{-.4em}
In the domain of medical imaging where severe class imbalance is the norm, the F1 score often is the metric of choice instead of accuracy. To test the generalization ability of solutions, we calculated the F1 score over all the images in Testset-1. For Testset-2, to measure personalization: how well a model performs on individual patients, patient-wise F1 scores were calculated over images from the same patient and these scores were averaged over all patients in the test dataset. 

\section{EXPERIMENTAL SETUP}
\vspace{-.6em}
\subsection{Data Augmentation}
\label{ssec:augment}
\vspace{-.5em}
We used random greyscale transformation with $p=0.2$, color jitter with $p=0.8$, random resized crop with $scale=(0.7,1)$, random horizontal flip, and finally, normalization with a mean of 0.1706 and a standard deviation of 0.2112. We found 0.7 to be the optimal scale for random resized crop while keeping other augmentations constant.

\vspace{-.7em}
\subsection{5-fold Cross Validation}
\label{ssec:implem}
\vspace{-.5em}
We performed a 5-fold cross-validation where we partitioned the data into 5 folds with 80\% in the train set and 20\% on the validation set. On these 5 different folds, we trained our models, ran inference on the test set after every epoch, and averaged the confidence scores to obtain the final binary decision for each biomarker. 

\vspace{-1em}
\subsection{Code Environment and Setup}
\label{ssec:hardware}
\vspace{-.5em}

For convolution-based models implemented in Tensorflow, we used \href{https://www.kaggle.com/}{Kaggle} TPU VM v3-8 instances paired with 330GB RAM. Due to the limited support of state-of-the-art models on TPU, we mainly used this setup for pilot experiments. For transformer-based models (implemented in PyTorch 2.0.1\cite{NEURIPS2019_9015} and `timm' \cite{rw2019timm} library with the weights hosted on \href{https://www.huggingface.co/}{Hugging Face}), we used \href{https://www.kaggle.com/}{Kaggle} Nvidia P100 GPU instances with 16GB VRAM, 13GB RAM, and 19GB disk space. We used scikit-learn \cite{pedregosa2011scikit} libraries for other auxiliary needs. The runtime of our complete MaxViT pipeline, including training, validation, and inference, was approximately 11 hours, while that of our EVA-02 pipeline was approximately 7 hours.

\vspace{-1em}
\subsection{Hyperparameters}
\label{ssec:hyperparam}
\vspace{-.3em}
We used AdamW\cite{adamw} optimizer with default initialization and set the initial learning rate to \(3 \times 10^{-5}\). We used the Exponential Learning Rate Scheduler, with a weight decay of 0.9. For convolution-based models, we used 128 as the batch size and trained models for 35 epochs, with early stopping based on the best cross-validation F1 score. For transformer-based models, we used the effective batch sizes 8 for MaxViT and 16 for both EVA-02 and ViT. We trained all vision transformer models for two epochs. We found all ViT models overfit the training data after 2 epochs.

\vspace{-.6em}
\section{RESULTS AND DISCUSSIONS}
\label{sec:page}\label{sec:results}
\vspace{-.5em}

\subsection{Baselines}
\label{ssec:baselines}
\vspace{-.3em}
To establish a baseline, we trained multiple variants of ResNet \cite{he2016deep} models and Inception \cite{szegedy2015going} models. We find that model size or ImageNet performance \cite{deng2009imagenet} are not reliable indicators of its suitability for the task at hand (Table \ref{tab:model_comparison}). InceptionResnetV2\cite{szegedy2017inception} (55.84 M parameters) proved to be the most effective model with an F1 score of 0.686 and the much smaller InceptionV3 (23.83 M parameters) model performed comparably with an F1 score of 0.682 (Table \ref{tab:model_comparison}).

\begin{table}[ht]
    \centering
    \begin{tabular}{lccc}
        \toprule
        Model & Param(M) & ImageNet & Test F1 \\
        \midrule
        ConvNextBase & 88.59 & \textbf{87.13} & 0.612 \\
        Resnet50 & 25.57 & 75.30 & 0.634 \\
        Resnet152 & 66.84 & 78.57 & 0.649 \\
        Resnet101 & 44.57 & 78.25 & 0.657 \\
        EfficientNetV2L & 118.52 & 86.80 & 0.662 \\
        InceptionV3 & 23.83 & 78.95 & 0.682 \\
        InceptionResnetV2 & 55.84 & 80.46 & \textbf{0.686} \\
        \bottomrule
    \end{tabular}
    \caption{Comparison of Convolution-based Models. We report the number of model parameters, Top1 Accuracy on the ImageNet \cite{deng2009imagenet} dataset collected from \href{https://paperswithcode.com/sota/image-classification-on-imagenet}{PapersWithCode}, and F1 score on Testset-1. All models were evaluated using 5-fold cross-validation.}
    \label{tab:model_comparison}
\end{table}

\vspace{-.5em}
We note that, 5-fold cross-validation boosts Testset-1 scores substantially. Initial experiments revealed that our best-performing convolution-based model, InceptionResnetV2 consistently scored ~0.66 when trained on random 80\% splits of the train set. However, using cross-validation, InceptionResnetV2 consistently scored around ~0.68. As such, we used cross-validation in all further experiments. 
Individually, MaxViT and EVA-02 models scored ~0.68 while with cross-validation they scored ~0.71.

\vspace{-.9em}
\subsection{Ablation Study with CBAM}
\vspace{-.2em}
Our ablation study involving the addition of CBAM \cite{CBAM} to InceptionResnetV2 showed a substantial boost in F1 score from 0.686 to 0.696 (Table \ref{tab:irv2_comparison}) for a negligible increase in the network complexity (i.e., parameter count increased by only \~0.37\%; not reported in the table). Notably, this boost in performance inspired us to move to vision transformer models. 

To understand the reason for the improved F1 scores, we calculated the F1 score across biomarker types individually and discovered that CBAM improved the performance on certain biomarkers substantially while showing marginal improvement in others. It even registered a deterioration, albeit only slightly, in one case. Therefore, we hypothesize that the attention module improved the detection of local biomarkers.

\begin{table}[ht]
    \centering
    \begin{tabular}{llccc}
        \toprule
        Biomarker & Type & IRV2 & IRV2\_CBAM & VIT\_BASE \\
        \midrule
        IRHRF & L & 0.709 &  {0.746} \textbf{(+)} & 0.773 \textbf{(+)} \\
        PAVF &  G & 0.610 & 0.609 & 0.662 \textbf{(+)}\\
        FAVF &  G & 0.837 & 0.841 & 0.869 \textbf{(+)}\\
        IRF &  L & 0.557 & {0.599} \textbf{(+)} & 0.552 \\
        DRT/DME & L/G &  0.599 & {0.628} \textbf{(+)} & 0.594\\
        VD &  G & 0.753 & 0.759 & 0.755\\
        \midrule
        Overall & & 0.686 & 0.696 & 0.701\\
        \bottomrule
    \end{tabular}
    \caption{Comparison of InceptionResnetV2 with (IRV2\_CBAM) and without (IRV2) CBAM. L (G) in the type column refers to Local (Global). For individual biomarker types, a plus sign in the bracket beside a score indicates significant improvement against the score of the network to its immediate left column. All models were evaluated using 5-fold cross-validation.}
    \label{tab:irv2_comparison}
\end{table}

\vspace{-.6em}
Although adding an attention mechanism in the form of CBAM to InceptionResnet specifically improves the performance on local biomarkers, we find no such correlation when comparing convolution-based models and the purely attention-based ViT \cite{dosovitskiy2020image} architectures. This suggests the need for explicit convolution in addition to attention for optimal biomarker detection.

\vspace{-.6em}
\subsection{Efficacy of combining  Convolution and Attention}
\vspace{-.2em}

MaxViT\cite{tu2022maxvit} is a vision transformer model composed of multiple MaxViT blocks where each block performs convolution, strided/block attention, and dilated/grid attention. The addition of explicit convolution makes MaxViT ideal for biomarker detection. We achieved an F1 score of 0.718 (Table \ref{tab:biomarker_comparison}) using the base variant of the MaxViT model, which is a substantial improvement over IRV2\_CBAM and ViT\_BASE. However, MaxViT does not utilize global attention across all image tokens, which motivated us to test EVA-02 \cite{EVA02}, a plain Vision Transformer model that improves upon the standard ViT \cite{dosovitskiy2020image} by using a 1B parameter EVA-CLIP model as its teacher. The parameter counts of MaxViT and EVA-02 are 119.88M and 87.12M respectively. Comparing MaxViT and EVA-02 across the 6 biomarkers,  we see that EVA-02 performs noticeably better on global biomarkers despite being smaller of the two. We hypothesize that MaxViT's sparse attention improves local biomarker detection while EVA-02's true attention excels at detecting global features.

\vspace{-1em}
\subsection{Ensembling Results}
\vspace{-.4em}
While our simple ensembling does boost the test set F1 score to 0.720 (not shown in the table for brevity), the finer-grained ensembling scheme yields an even greater performance with an improved F1 score of 0.724.
\vspace{-.4em}

\begin{table}[ht]
    \centering
    \begin{tabular}{llccc}
        \toprule
        Biomarker & Type & MaxViT & EVA-02 & Ensemble \\
        \midrule
        IRHRF & L & 0.774 & 0.731 & \textbf{0.779} \\
        PAVF & G & 0.677 & \textbf{0.701} & 0.688 \\
        FAVF & G & 0.868 & 0.874 & \textbf{0.879} \\
        IRF & L & \textbf{0.611} & 0.575 & 0.600 \\
        \small{DRT/DME} & L/G & 0.615 & 0.593 & \textbf{0.618} \\
        VD & L & 0.764 & 0.779 & \textbf{0.782} \\
        \midrule
        Overall & - & 0.718 & 0.709 & \textbf{0.724} \\
        \bottomrule
    \end{tabular}
    \vspace{-.2em}
    \caption{F1 Score comparison of MaxViT, EVA-02, and their ensemble across various biomarkers on the validation set. The models have been ensembled by averaging their output probabilities.(L: Local, G: Global)}
\label{tab:biomarker_comparison}
\end{table}

\vspace{-.7em}

Our MaxViT-EVA02 ensemble pipeline achieved a patient-wise F1 score of 0.814 in Testset-1 and 0.8527 in Testset-2 -- 3.8\% higher than the next best solution  (\footnote{https://alregib.ece.gatech.edu/2023-vip-cup/}{leaderboard}). 
\vspace{-1.0em}
\subsection{Leveraging Unlabeled Training Data}
\vspace{-.4em}
We initially explored contrastive learning \cite{kokilepersaud2023clinically} with Inception-based models but were unable to reproduce the reported gains, and Inception-ResNetV2 performed no better than the fine-tuning baseline. Predicting all eight labels (six biomarkers and two clinical labels) also failed to improve performance. Attempts at \textbf{pseudo-labeling}, where high-confidence predictions ($>0.95$) from a fine-tuned Inception-ResNetV2 model were used to augment the dataset, resulted in significant performance deterioration ($F1 = 0.519$). Similarly, experiments with I-JEPA \cite{assran2023selfsupervisedlearningimagesjointembedding}, an unsupervised pretraining method, led to further performance declines, suggesting this methodology was not well-suited for our specific task.

We believe our initial attempt with pseudo-labeling lacked a strong baseline model. As we now use predictions for a total of 10 models (5-fold MaxViT and 5-fold EVA02) for labeling, we get higher-quality pseudo-labels. We performed an ablation study (Table \ref{tab:pseudolable_ablation}) to assess the impact of different combinations of pseudo-label pretraining and fine-tuning.

\begin{table}[ht]
    \centering

    \begin{tabular}{lcccc}
    \toprule
    \small{Biomarker} & \small{MaxViT$_{p}$} & \small{MaxViT$_{f}$} & \small{MaxViT$_e$} & \small{MaxViT$_{pf}$} \\
    \midrule
    IRHRF & 0.475 & 0.748 & 0.774 & \textbf{0.783}\\
    PAVF & 0.479 & 0.662 & 0.\textbf{677} & 0.655\\
    FAVF  & 0.723 & 0.846 & \textbf{0.868} & 0.865\\
    IRF  & 0.304 & 0.607 & 0.611 & \textbf{0.632}\\
    \small{DRT/DME}  & \textbf{0.719} & 0.581 & 0.615 & 0.642\\
    VD  & 0.198 & 0.755 & 0.764 & \textbf{0.771}\\
    \midrule
    Overall & 0.375 & 0.700 & 0.718 & \textbf{0.725}\\
    \bottomrule
    \end{tabular}

    \caption{F1-Score comparison of MaxViT$_p$ (only pretrained on pseudo-labeled data), MaxViT$_f$ (only fine-tuned on labeled training data), MaxViT$_e$ (5-model ensembled MaxViT) and MaxViT$_{pf}$ (pseudo-label pretrained before fine-tuning) for various biomarkers on the validation set. We considered only one model per type for the $p$, $f$ and $pf$ variants.}
    \label{tab:pseudolable_ablation}
    \vspace{-1.5em}
\end{table}

This knowledge distillation from our larger ensemble model enabled a single MaxViT to slightly outperform our MaxViT-EVA02, while requiring only a fraction of the inference time and computational resources. Incorporating this distilled MaxViT into our original pipeline would undoubtedly yield further performance gains, but we leave this exploration for future work. 

\vspace{-.8em}
\subsection{Complexity-Performance tradeoff:}
\begin{figure}[htbp]
    \centering
    \includegraphics[width=.9\linewidth]{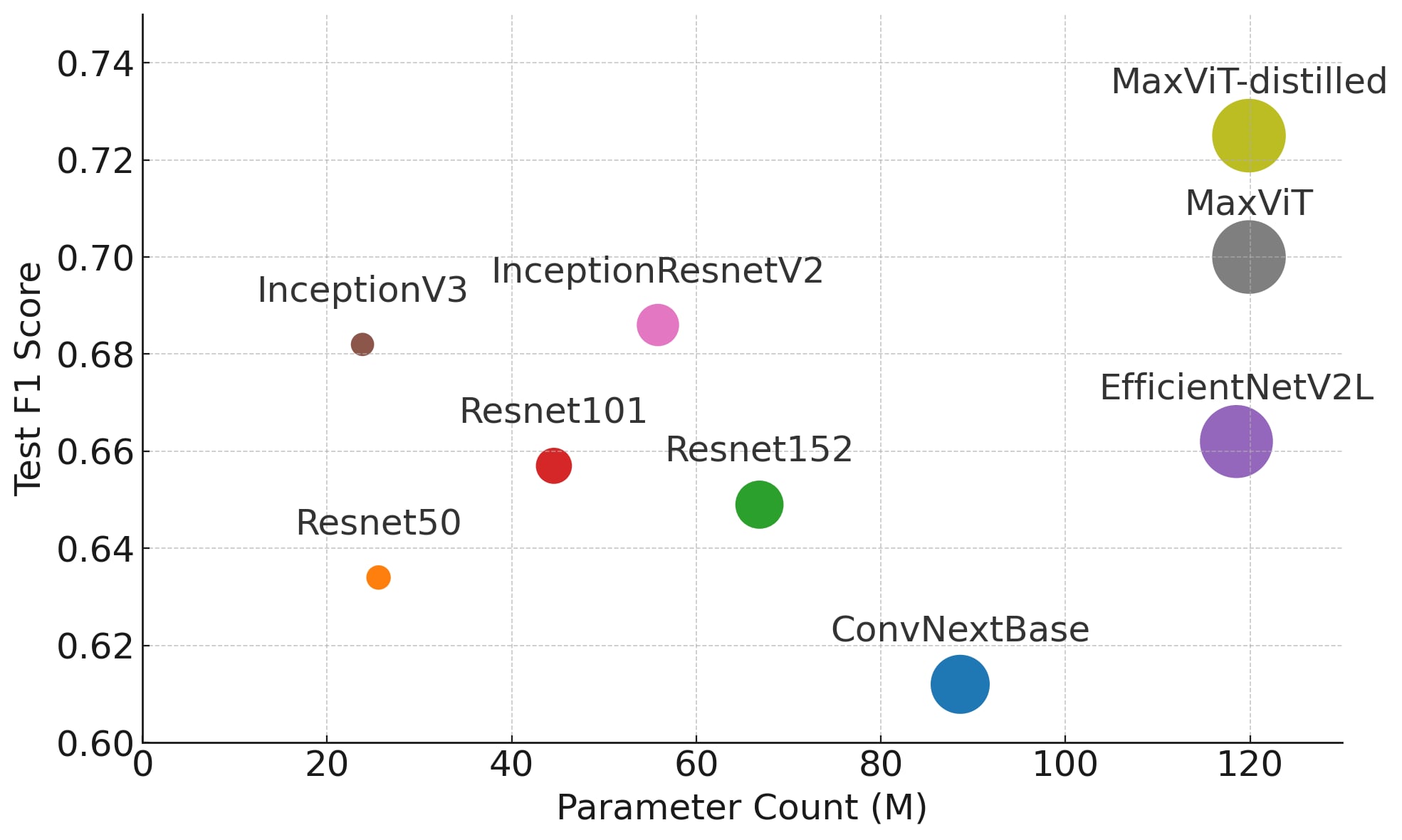}
    \caption{Tradeoff between model complexity (parameter count in millions) and performance (Test F1 Score) for various vision architectures.}
    \label{fig:PvC}
    \vspace{-.8em}
\end{figure}

\vspace{-.4em}
Model size alone is not a reliable indicator of performance in OCT biomarker detection as highlighted by the performances of the Inception, ResNet, ConvNext, and EfficientNet model variants (Fig. \ref{fig:PvC}). Notably, our distilled MaxViT slightly outperforms the MaxViT and EVA02 ensembles ( Table \ref{tab:biomarker_comparison}, not shown in the figure), which have roughly an order of magnitude more parameters. 
\vspace{-.5em}
\vspace{-.8em}
\subsection{Analysis of Outlying Patient-wise F1 Scores}
\vspace{-.3em}
In the analysis of cases where the model exhibited a low F1 score in detecting biomarkers from OCT scans, several patterns were observed. Patient 01-002 at week 40 and patient 02-044 at week 0 presented with severe spots, resulting in F1 scores of 0.64 and 0.55, respectively. Moderate spots were identified as the likely cause for the low F1 scores of 0.6 in patients 01-007 at week 100 and 01-049 at week 0. Additionally, patient 01-043 at week 100 exhibited a severe artifact, leading to the lowest F1 score of 0.37. Moderate artifacts were also noted in patients 01-049 and 02-044 at week 100, with F1 scores of 0.6 and 0.52, respectively. However, the likely cause for the low F1 scores observed in patients 01-019, 01-036, and 01-054 at week 100 (F1 scores of 0.51, 0.62, and 0.48) are not immediately evident to non-medical professionals. We leave a more thorough analysis and subsequent pipeline adjustments as future work.
\vspace{-1em}
\section{CONCLUSION}
\label{sec:conclusion}
\vspace{-.8em}

In this work, we outlined the methodology for our study on Ophthalmic Biomarker Detection and presented the underlying motivations for pipeline design decisions. Our findings indicate that Vision Transformer (ViT) models have begun to consistently outperform their Convolutional Neural Network (CNN) counterparts. Additionally, we observed that k-fold cross-validation and model ensembling are effective techniques for leveraging the entire dataset and improving generalization. Finally, utilizing the abundance of unlabeled data through knowledge distillation proves to be an efficient approach for enhancing model performance. For future work, we plan to explore our pipeline's generalizability to patient data collected from diverse sources, and interpretability analysis to improve trustworthiness.

\vspace{-.8em}
\section{ACKNOWLEDGEMENT}
\label{Ack}
\vspace{-.6em}
We would like to extend our sincere gratitude to Dr. S.M. Rezwan Hussain, a distinguished ophthalmologist at the Eye Department, Combined Military Hospital (CMH), Dhaka, Bangladesh, for his invaluable insights and expertise regarding biomarker classification according to their spatial extent.

\vspace{-1em}
\bibliographystyle{IEEEbib}
\bibliography{refs}

\end{document}